\documentclass[twocolumn, amsmath,amssymb, superscriptaddress]{revtex4}

\usepackage{dcolumn}
\usepackage{bm}
\usepackage{amsmath}
\usepackage{amsfonts}
\usepackage{graphics}
\usepackage[dvips]{graphicx}
\usepackage{times}
\usepackage{setspace}

\begin{document}

\title{Bankruptcy Risk  Model  and Empirical Tests}
\author{Boris~Podobnik}
\affiliation{Center for Polymer Studies and Department of Physics, Boston University, Boston, Massachusetts 02215, USA}
\affiliation{Faculty of Civil Engineering, University of Rijeka, 51000 Rijeka, Croatia}
\affiliation{Faculty of Economics, University of Ljubljana, 1000 Ljubljana, Slovenia}
\author{Davor Horvatic}
\affiliation{Faculty of Science, University of Zagreb, 10000 Zagreb, Croatia}
\author{Alexander M. Petersen}
\affiliation{Center for Polymer Studies and Department of Physics, Boston University, Boston, Massachusetts 02215, USA}
\author{Branko Uro\v{s}evi\'{c}}
\affiliation{Faculty of Economics, University of Belgrade, Serbia}
\author{H. Eugene Stanley}
\affiliation{Center for Polymer Studies and Department of Physics, Boston University, Boston, Massachusetts 02215, USA}
\date{\today}

\begin{abstract}
We analyze the size-dependence and temporal stability of firm bankruptcy
risk in the US economy by applying Zipf scaling techniques. We focus on a
single risk factor---the debt-to-asset ratio $R$---in order to study the
stability of the Zipf distribution of $R$ over time. We find that the Zipf exponent Êincreases during market crashes, implying that firms go
bankrupt with larger values of $R$. Based on the Zipf analysis, we employ
Bayes' theorem and relate the conditional probability that a bankrupt firm
has a ratio $R$ with the conditional probability of bankruptcy for a firm
with a given $R$ value. For 2,737 bankrupt firms, we demonstrate
size-dependence in assets change during the bankruptcy proceedings. Pre-petition firm assets and
petition firm assets Êfollow Zipf distributions but with different
exponents, meaning that firms with smaller assets adjust their assets more
than firms with larger assets during the bankruptcy process. We compare bankrupt firms with
non-bankrupt firms by analyzing the assets and liabilities of two
large subsets of the US economy: 2,545 Nasdaq members and 1,680 NYSE
members. We find that both  assets and liabilities Êfollow a Pareto
distribution. This is not a trivial
consequence of the Zipf scaling Êrelationship of firm size quantified
by employees---while the market capitalization of Nasdaq stocks follows
a Pareto distribution, this is not true for NYSE stocks. We propose a
coupled Simon model that simultaneously evolves both assets and debt with
the possibility of bankruptcy, and we also consider the possibility of firm mergers.
\end{abstract}

\maketitle

Complex systems are commonly coupled together and therefore should be
considered and modeled as interdependent. It is important to study the
conditions of interaction which may lead to mutual failure, the
indicators of such failure, and the behavior of the indicators in times
of crisis.  As an indicator of economic failure, default risk is defined
as the probability that a borrower cannot meet his or her financial
obligations, i.e. cannot make principal and/or interest payments
\cite{Duffie,Lando}.  Accordingly, it is important to better understand
default risk
\cite{Duffie,Lando,Beaver66,Altman68,Altman77,Ohlson80,Scott,Zmijewski84,Schary,Dichev,Shumway,Beaver}
and its relation to firm growth
\cite{Simon1,Ijiri,Jovanovic1,Sutton1,Sutton2,pnas1}, and how they
behave in times of crisis.

In describing both natural and social phenomena, size-dependent scaling
is an essential technique for understanding the common relations between
large and small scales.  One reason for the importance of scaling
relationships is that they capture salient, robust features of a system.
Sala-i-Martin found by analyzing the scaling relation between GDP growth
rates and initial GDP over the thirty-year period 1960--1990 that poor
countries grow slower than rich countries
\cite{martin96}. Size-dependent scaling of the standard deviations of
firm growth rates and country growth rates are reported in
\cite{Stanley3} and \cite{Lee}.  In finance, Fama and French
demonstrated that market risk depends upon the firm size \cite{fama}.

We find that book values of assets and debt of the U.S. companies that
filed for bankruptcy in the past twenty years follow a Zipf scaling
(power-law) distribution. The same is true for the values of assets and
debt of non-bankrupt firms comprising the Nasdaq.  We focus our
attention on a single risk indicator, the debt-to-asset ratio $R$, in
order to analyze stability of the scaling exponent or establish
crossover regions.  In order to capture Pareto and Zipf laws, the
literature has typically focused on a single Simon model
\cite{Simon1,Ijiri,Sutton1,Sutton2,pnas1} describing a single dynamic
system which does not interact with others.  We model the growth of debt
and asset values using two dependent (coupled) Simon models with two
parameters only, bankruptcy rate and another parameter controlling
debt-to-asset ratio.  The Zipf law scaling predictions of the coupled
Simon model are consistent with our empirical findings.

\section{Data Analyzed}

Our dataset consists of medium-size and large U.S. companies that filed
for bankruptcy protection in the period 1990--2009.  We obtain our data
from the URL {\sf www.BankruptcyData.com}, one of the most comprehensive
bankruptcy dataset currently available on the web.  There is also a
bankruptcy data set available at {\sf
  http://bdp.law.harvard.edu/fellows.cfm}, but with smaller firms and no
debt data. Our dataset includes data on 2,737 public and private
firms. The book value of firm assets in the database ranges from 50
million to almost 700 billion USD.

\begin{itemize}

\item[{A.}] For each firm in our sample we know the pre-petition book
  value of firm assets $A_{a}$ and the effective date of bankruptcy.
  From the court petition documents we find the petition book value of
  firm assets $A_{b}$, as well as book value of total debt, $D_{b}$. As
  an example Lehman Brothers filled a petition on September 15, 2008,
  listing the debt $D_{b}$ and assets $A_{b}$ on May 31, 2008.  Thus,
  $A_{b}$, $A_{a}$, and $D_{b}$ quantify the debtor's condition before
  declaring bankruptcy.  We are able to obtain $A_{b}$ and debt $D_{b}$
  for 462 firms. Note that Refs.~\cite{Altman77}, \cite{Ohlson80}, and
  \cite{Beaver} studied 53, 105, and 585 bankrupt firms, respectively.
  There is often a substantial change in the debt and assets of a
  company in the time period preceding bankruptcy.  Hence for each firm
  we calculate the debt-to-assets (leverage ratio)
\begin{equation}
   R \equiv  D_b/A_b,
\label{ratio}
\end{equation}
from the total debt $D_b$ and assets $A_b$ estimated simultaneously.
Note that economics has a parallel treatment, known as Tobin's Q theory
of investment which also focuses on a single factor, Q \cite{Tobin}.
 
\qquad In the literature on ratio analysis
\cite{Altman68,Ohlson80,Zmijewski84}, multiple financial ratios are used
for predicting probability of default, such as the ratio of total
liabilities to total assets.  Adding more factors would likely improve
the predictive power of the model, so we consider only one risk factor,
namely the debt-to-assets ratio $R$ which captures the level of company
indebtedness. We do this for two reasons: (a) to make a model as simple
as possible, and (b) in order to simplify our study regarding whether
market crashes and global recessions affect the scaling existing in
bankruptcy data.  In order to relate the probability of bankruptcy to
$R$, we analyze the scaling relations that quantify the probability
distribution of firms that entered into bankruptcy proceedings with
particular values of $A_{b}$ and $R$.  Our analysis includes very few number
 of  young start-up firms, for which the age of the firm also factors into the
probability of bankruptcy in addition to $R$.  In 2009 we find that the
average lifetime of the 215 bankrupt firms analyzed was $35.8 \pm 33.9$
years, and the minimum lifetime was 3 years.

\item[{B.}] We analyze market capitalization, assets and liabilities of
  2,545 firms traded on the Nasdaq over the three-year period from 2006
  to 2008. We also analyze assets and liabilities of 1,680 firms traded
  on the New York Stock Exchange (NYSE) in the period from 2007 to
  2009. Also, we analyze market capitalization of NYSE members over the
  period 2002-2007.

\end{itemize}

\section{Quantitative Methods}

Our analysis is closely related to the literature on firm size \cite{EL,
  Axtell1}.  Analyzing data from the U.S. Census Bureau,
Ref.~\cite{Axtell1} reported that firm sizes of the U.S. firms follow a
Zipf law: the number of firms larger than size $s$ is $s^{-\zeta}$,
where $\zeta\simeq 1$.  The Zipf distribution is found for the
distribution of city sizes \cite{Gabaix1} and the distribution of firm
sizes \cite{Axtell1,Gabaix09}.

The cumulative distribution is a simple transformation of the Zipf
rank-frequency relation, where the observations $x_i$ are ordered
according to rank $r$ from largest ($r \equiv 1$) to smallest.  For
Pareto-distributed variables $ s $ with cumulative distribution $P(s >
x) \sim x ^{-\zeta'}$, the Zipf plot of size $s$ versus rank $r$
exhibits a power-law scaling regime with the scaling exponent $\zeta$,
where
\begin{equation}
\zeta = 1 / \zeta'.
\label{cumulative}
\end{equation}

\section{Results of Analysis}

\begin{figure}
\centering{\includegraphics[width=9cm,angle=-0]{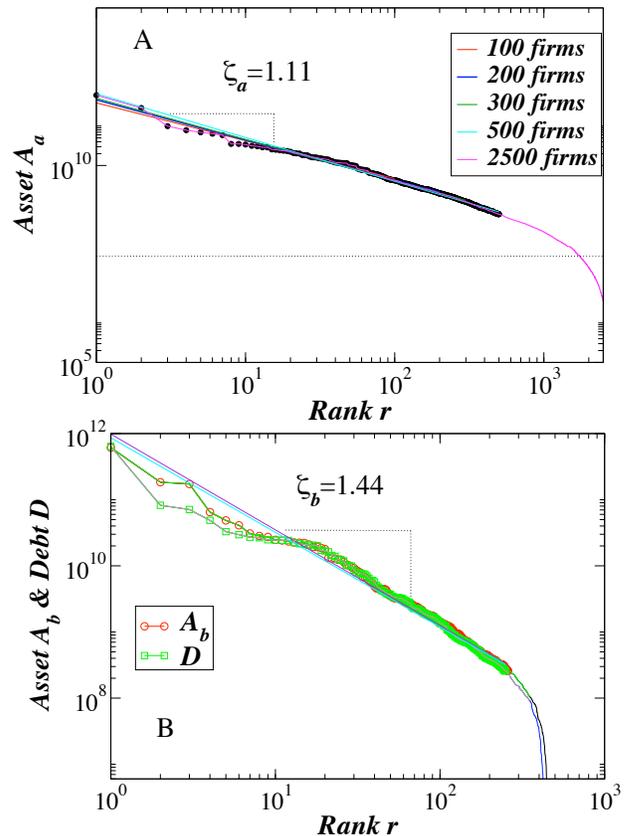}}
\caption{Zipf plot of U.S. bankrupted firm assets.  (a) Zipf plot
  calculated for firms over the last 20 years between pre-petition total
  assets, $A_a$ versus rank.  Deviation from the Zipf law is due to the
  fact that the data set includes mainly the firms with assets larger
  than 50 million dollars (dotted line). (b) Zipf plot of
  U.S. bankrupted firms of debt versus rank---462 firms in total---along
  with Zipf plot of book value asset and rank. The two plots practically
  overlap.}
\label{fig.1}
\end{figure}

\begin{figure}
\centering{\includegraphics[width=9cm,angle=-0]{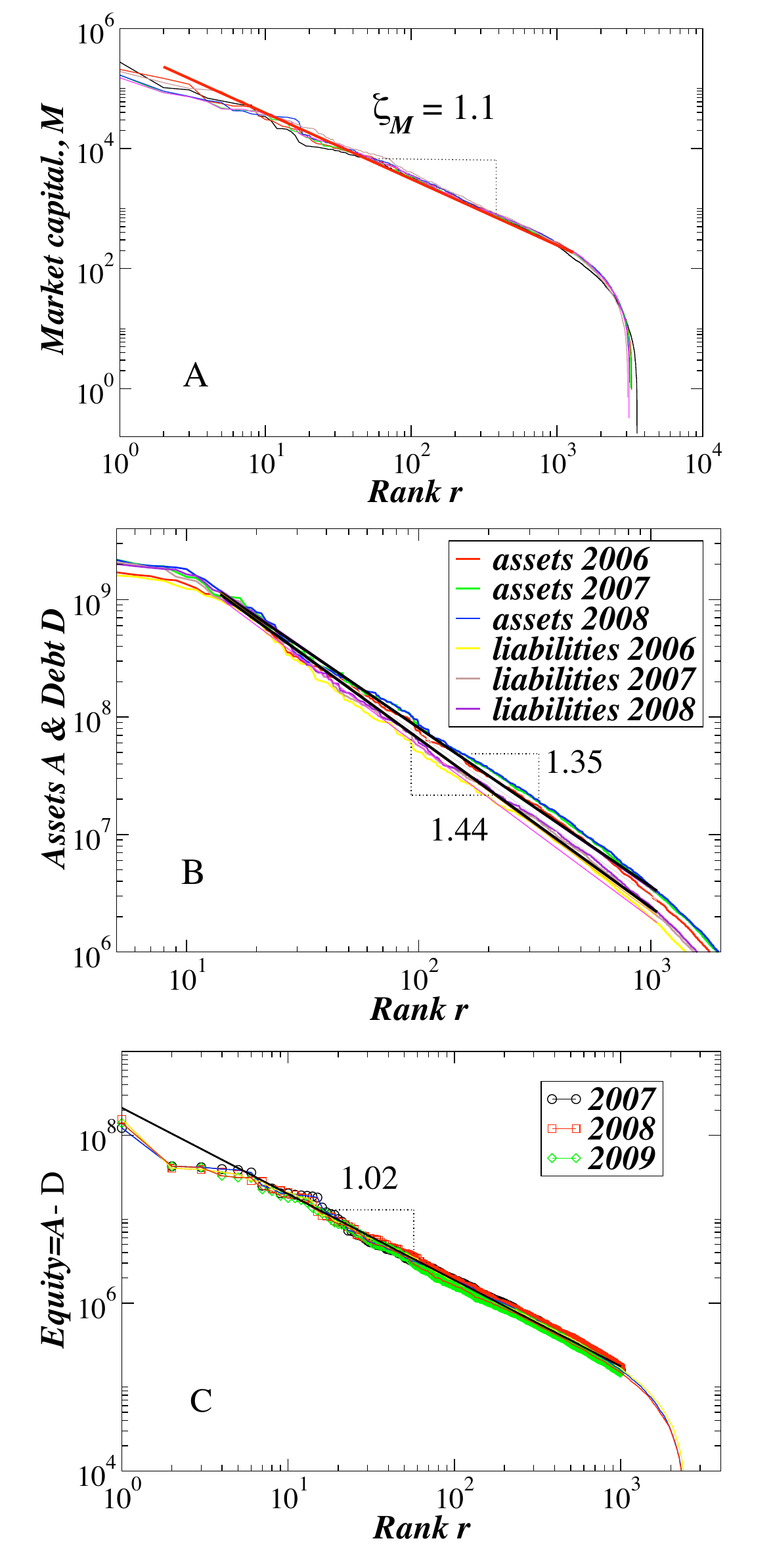}}
\caption{(a) Zipf plot of market capitalization $M$ versus rank $r$ for
  the Nasdaq members for each year of 6 years. We find practically the
  same Zipf law for the largest 1,000 companies as we find for the
  assets $A_{a}$ of bankrupted firms in Fig.~1(b). (b) For the Nasdaq
  firms both assets and liabilities follow a Zipf plot. (c) Book value
  of equity of stock traded at Nasdaq, defined as assets less
  liabilities, follows a Zipf law.}
\label{fig.2}
\end{figure}

Figure~1(a) shows the Zipf plot for pre-petition book value of assets
$A_a$. The data are approximately linear in a log-log plot with the
exponent
\begin{equation}
\zeta_{{a}} = 1.11 \pm 0.01,
\label{market}
\end{equation}
obtained using the OLS regression method.  For the U.S. data on firm
size (measured by the number of employees), Ref.~\cite{Axtell1} reported
the value $\zeta \approx 1$.  Hence, prior to filing for bankruptcy
protection, the book value of firm assets for companies that later
underwent bankruptcy satisfies a scaling relation similar to that in
\cite{Axtell1}.  The firms with a rank larger than $\approx 500$ start
to deviate from the Zipf law, a result of finite size effects as found
in data on firm size \cite{Axtell1}.

It is known that the market equity of firms that are close to bankruptcy
is typically discounted by traders \cite{Dichev,Beaver}.  In order to
study if those changes are size-dependent during the time of bankruptcy,
we test whether there is a difference in scaling behavior between
pre-petition and petition firm assets.  Figure~1 {\it B} ranks the firm
book value of assets $A_{b}$ and firm debt $D_b$.  We find
\begin{equation}
         \zeta_{{b}} = \zeta_{D} = 1.44 \pm 0.01.
\label{book}
\end{equation}
Also, a Zipf law is found for the distribution of total liabilities of
bankrupted firms in Japan \cite{Hideki,Fujiwara}.

We obtain that $\zeta_{b}>\zeta_{a}$, a discrepancy that could be of
potential practical interest.  To clarify this point, if $A_{b}$ is
related to $A_{a}$ by a constant $A_{b}/A_{a} \equiv c$, we would
observe $\zeta_{{a}} = \zeta_{{b}}$.  However, we observe an increasing
relation $A_{a}/A_{b} \propto r^{\zeta_{{b}} -\zeta_{{a}}}$ with rank
$r$, meaning that bankrupt firms with smaller $A_{a}$ have larger
relative adjustments than do bankrupt firms with larger $A_{b}$. By
using a method proposed in Ref. \cite{GabaixJBES} we obtain $\zeta_b=
1.39$, close to the exponent we found in Eq.~(\ref{book}).

\begin{figure}
\centering{\includegraphics[width=9cm,angle=-0]{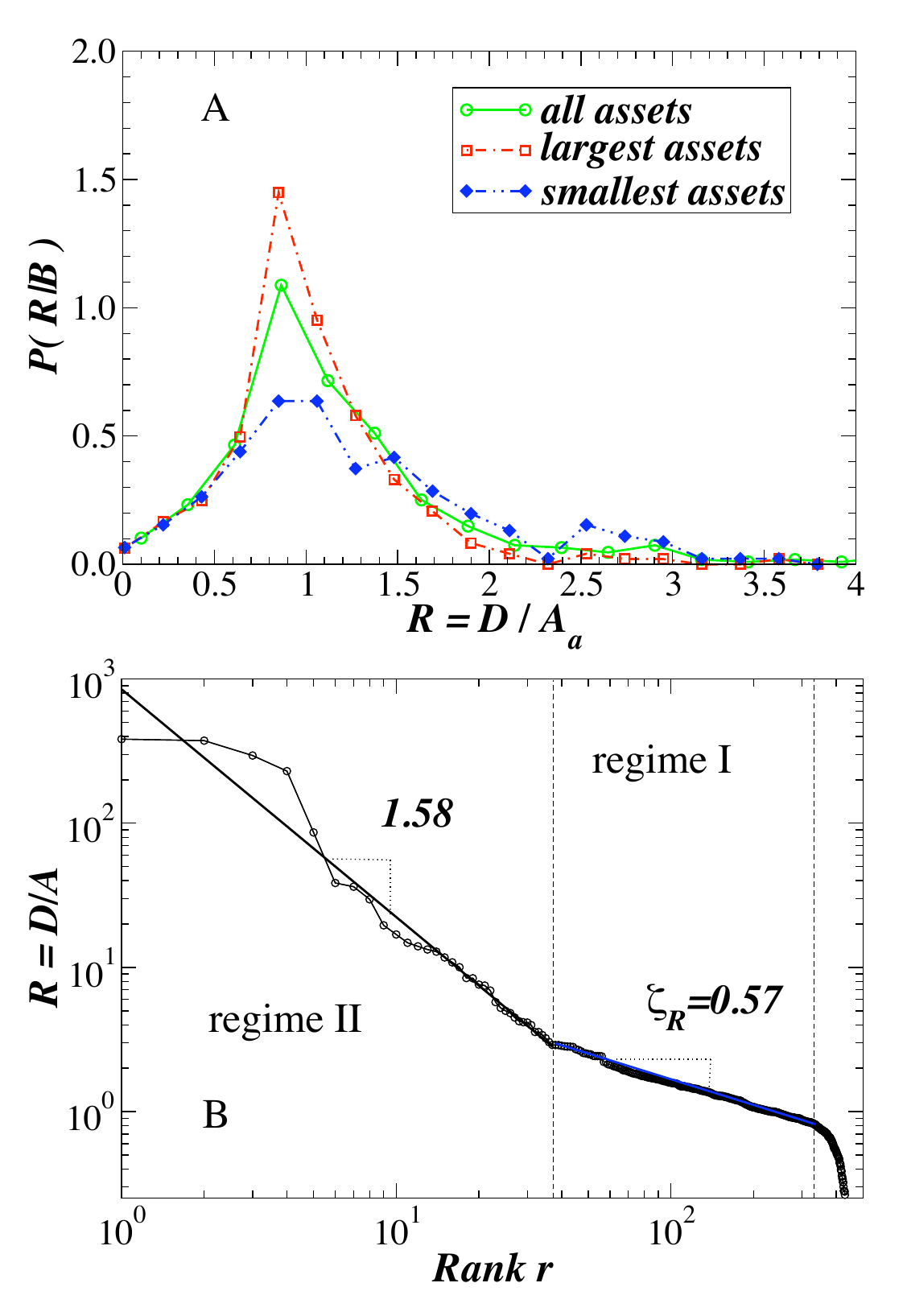}}
\caption{Bankruptcy risk based on petition book value asset $A_{b}$ and
  debt $D_b$.  (a) We find the distribution of $R=D_b/A_{b}$ for the
  bankrupted firms. One may calculate the probability that a firm with a
  given ratio $R$ will go bankrupt when its ratio is $\leq R$.  (b) For
  the ratio values $3 > R > 0.8$ (67\% of all data), we show the Zipf
  plot that can be approximated by a Pareto distribution with $ \zeta_R
  = 0.57$. The same regime we fit with the power-law tail of pdf and
  obtain $0.79 R^{-2.72}$ where the exponent $\zeta'+1=2.73$ is in
  agreement (see Eq.~(\ref{cumulative})) with the Zipf exponent
  $\zeta=0.57$.  For the largest ratio values $ R > 3$ (7\% of all
  data), we find a crossover to a power-law regime with $\zeta_R =
  1.58$.}
\label{fig.3}
\end{figure}

Our analysis of bankruptcy probability is, due to data limitation, based
on book values.  One may argue that a more relevant analysis would be
based on market values of assets and liabilities.  We now demonstrate
that using market instead of book values may in fact lead to similar
results. For this purpose, let us consider companies for which we have
both market and book value data, namely stocks that comprise the Nasdaq.
We begin by finding market capitalization of Nasdaq members for each
year from 2002 to 2007. The data are available at {\sf
  www.bloomberg.com}. Figure~2(a) shows the Zipf plot for market
capitalization deflated to 2002 dollar values. We find that the market
capitalization versus rank for the largest $\approx 1,000$ companies is
well described by a Zipf law with exponent $\zeta_{M}=1.1 \pm 0.02$, in
agreement with Ref.~\cite{Jovanovic3}.

In Fig.~2(b) we repeat the Zipf analysis using, this time, book values
of both assets and debt for the same Nasdaq stocks.  The scaling
exponents we observe in Fig.~2(b) are larger than the exponent observed
in Fig.~2(a).  However, market capitalization is best compared with book
value of equity $E\equiv A-D$, rather than assets $A$.  In Fig.~2(c) we
find that $E$ also exhibits Zipf scaling with exponent $\zeta_{E} = 1.02
\pm 0.01$, which is more similar to $\zeta_{M}$.  Therefore, we find
qualitatively similar scaling for the existing Nasdaq companies and for
companies before they entered into bankruptcy proceedings.

The probability of bankruptcy $P(R)$ is a natural proxy for firm
distress \cite{Dichev}. Previous studies analyzed defaults of firms
traded at NYSE, AMEX, and Nasdaq \cite{Dichev}. In contrast, the
majority of firms in our dataset are privately held companies.  For
bankrupt firms in Fig.~3(a) we show $P(R|B)$ for values of the
debt-to-assets ratio $0 < R < 4$. We truncate data to avoid outliers as
in Ref.~\cite{Shumway}.  We find $P(R|B)$ is right-skewed with a maximum
at $ R \approx 1$, and $\langle R \rangle = 1.4 \pm 1.5$.

Previous studies find that bankruptcy risk of NYSE and AMEX stocks is
negatively related to firm size \cite{Dichev}.  In order to test for
firm-size dependence of bankruptcy risk for mainly private firms using
{\sf BankruptcyData.com} with $R$ as bankruptcy measure, we divide the
$R$ values into two subsamples based on their value of $A_{b}$.  In
Fig.~3(a) we demonstrate qualitatively that $R$ is size-dependent.  The
pdfs for small $A_{b}$ and large $A_{b}$ are similar in that they both
show peaks at $R \approx 1$. However, firms with smaller assets, as
measured by $A_{b}$, have a larger probability of high debt-to-assets
ratios $R$ than firms with large assets $A_{b}$.

In addition, we test for the size-dependence by performing the
Mann-Whitney U-test, which quantifies the difference between the two
populations based on the difference between the asset ranks of the two
samples. (The null hypothesis is that the distributions are the same).
Since the test statistics U-value $=-5.60$, we reject the null
hypothesis thus confirming that $R$ depends on $A_{b}$ at the $p = 0.05$
confidence level.

In Fig.~3(b) we analyze the Zipf scaling for large $R$.  We find that
the Zipf plot can be approximated by two power-law regimes.  For
$\approx 300$ firms with $0.8 < R < 3$ (regime {\bf I}), we find a
power-law regime with $\zeta_{R} = 0.57 \pm 0.02$.  Hence, according to
Eq.~(\ref{cumulative}) we conclude that the cumulative distribution of
dangerously high $R$ values of bankrupt firms decreases faster with
$\zeta' \approx 1.72$ for large $R$ than the distribution of firm size
\cite{Axtell1} and firm assets with $\zeta' \approx 1$ (see Fig.~1). For
$R > 3$ (7\% of all data including predominantly financial firms), we
find that the Zipf plot exhibits a significant crossover behavior to a
power-law regime with $\zeta \approx 1.58$.

\begin{figure}
\centering{\includegraphics[width=9cm,angle=-0]{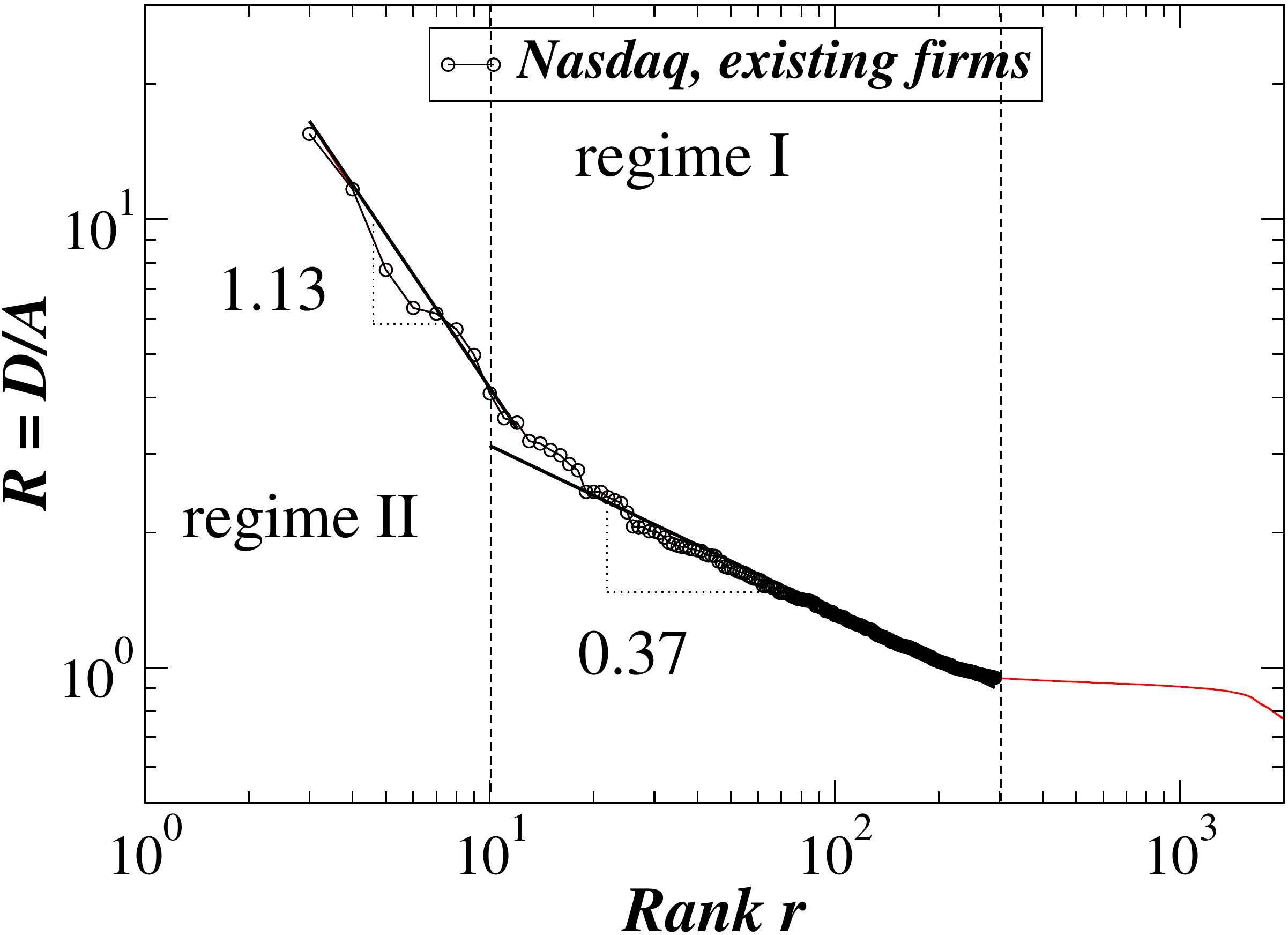}}
\caption{Zipf plot of debt-to-assets ratio $R$ and rank $r$ for the
  existing firms of the Nasdaq members over the last 3 years.  For the
  $\approx 300 $ ratio values smaller than 3.5 and larger than 0.95 the
  Zipf plot has exponent 0.37.  The same regime we fit with the
  power-law tail of pdf and obtain $1.54 R^{-3.6}$ where the exponent
  $\zeta'=3.6$ agrees [see Eq.~(\ref{cumulative})] with the Zipf
  exponent $\zeta=0.37$.}
\label{fig.4}
\end{figure}

The conditional probability $P(B|R)$ that an existing firm with
debt-to-assets ratio $R$ will file for bankruptcy protection may be of
significance to rating agencies, creditors, and investors.  According to
Bayes theorem, $P(B|R)$ depends on $P(R|B)$ (see Fig.~3), $P(B)$, the
probability of bankruptcy for existing firms, and $P(R)$ where as a
proxy for the existing companies we use companies constituting the
Nasdaq in the 3-year period between 2007--2009.  For this time period we
obtain book value of each firm's assets and liabilities (the latter
serving as a proxy for total debt). As a result we obtain 7,635 $R$
values with median value 0.48.  For existing Nasdaq members, Fig.~4
shows that the Zipf plot can be approximated by two power-law regimes,
where regime {\bf I} with $ 3.5 > R > 0.9$ yields $\zeta_e = 0.37 \pm
0.01$. Note that regime {\bf I} is similar to the one we find in
Fig.~3(b) for bankruptcy data. $P(B)$ may substantially change during
economic crises. Interestingly, Ref. \cite{DebtEPL} analyzes the debt-to-GDP (gross domestic product) ratio for
countries, in analogy to the debt-to-assets ratio for existing firms, and calculates a Zipf scaling exponent that  is approximately the same as the scaling exponent
calculated here for existing Nasdaq firms.

We estimate the scaling of $P(B|R)$ using Bayes' theorem,
\begin{eqnarray}
P(B|R) &=&\frac{P(R|B)P(B)}{P(R)} \approx 0.51~ P(B)
R^{1/\zeta_e-1/\zeta_R} \nonumber \\
&& \approx 0.51~ P(B)~ R^{0.95}
\label{bayes}
\end{eqnarray}
where we approximate $P(R|B)$ and $P(R)$ with power laws---$P(R|B)\sim
R^{-(1/\zeta_R+1)}\Delta R$ and $P(R)\sim R^{-(1/\zeta_e+1)}\Delta
R$. The value of the relevant exponents calculated for regime {\bf I}
are: $\zeta_{e} \approx 0.37$ (see Fig.~4) and $\zeta_{R}=0.57$ [see
  Fig.~3(b)], where $\zeta_R > \zeta_e$ implies that $P(B|R)$ increases
with firm indebtedness quantified by $R$.  The pre-factor 0.51
calculated for the regime {\bf I} we estimate from the corresponding
intercepts in pdfs [see Figs.~3(b) and 4].  In Fig.~4 we find a
pronounced crossover in the Zipf plot for very large values of the $R$
ratio.

\begin{figure}
\centering{\includegraphics[width=9cm,angle=-0]{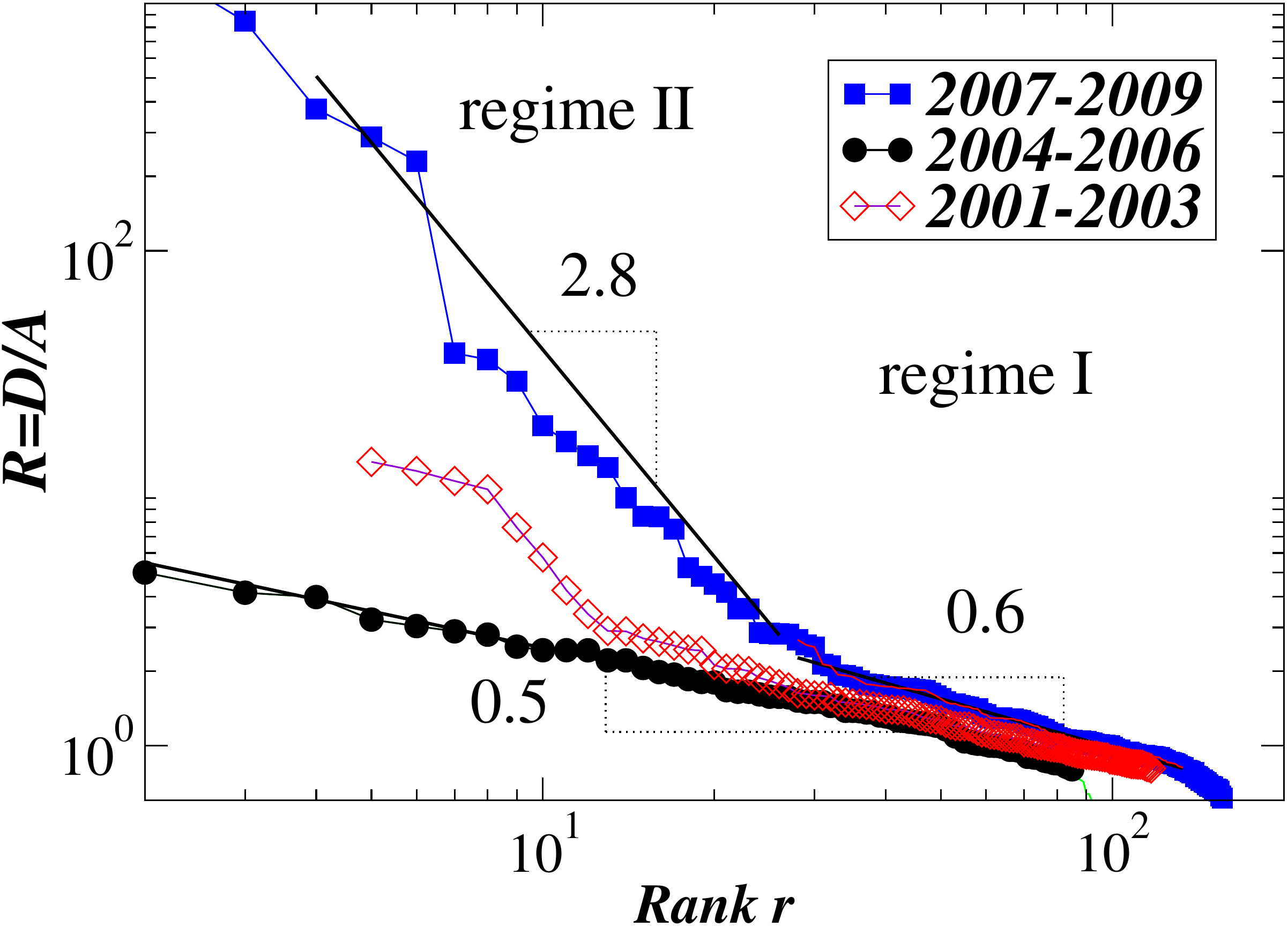}}
\vspace*{0.cm}
\caption{Zipf plot of debt-to-assets ratio $R$ versus rank $r$ for the
  bankrupt firms for the three different 3-year subperiods.  During the
  last three years characterized by recession the Zipf plot exhibits a
  crossover behavior. A smaller crossover in the Zipf plot also exists
  for the period 2001--2003 characterized by the dot-com bubble burst.}
\label{fig.5}
\end{figure}

In order to test whether market crash and global recession have
significant effects on the scaling we find in the bankruptcy data, in
Fig.~5 we analyze the Zipf scaling of the large $R$ values for three
different 3-year periods. For the period 2004--2006 we find a stable
Zipf plot characterized by an exponent $\zeta_{R} = 0.50 \pm 0.01$ close
to the value we found in Fig.~3(b) for all years analyzed.  For the
period 2001--2003 characterized by the dot-com bubble burst, we
find a less pronounced crossover in the Zipf plot between regime {\bf I}
with exponent $\zeta_{R} = 0.58 \pm 0.01$ and regime {\bf II}.  For the
period 2007--2009 we find that the Zipf plot exhibits a significant
crossover behavior between regime {\bf I} and regime {\bf II}.

Figure 5 demonstrates the existence of a relatively stable scaling
exponent (between 0.5 and 0.6) in regime {\bf I} over the 9-year period
2001-2009.  However, in times of economic crisis, e.g., the period
2007--2009, the exponent in regime {\bf I} increases, implying that
firms go bankrupt with larger values of $R$. According to
Eq.~(\ref{bayes}), in times of crisis $(\zeta_R \approx 0.6)$ $P(B|R)
\propto R^{1/\zeta_e-1/\zeta_R} ~\propto R^{1} $ shifts upward compared
to times of relative stability $(\zeta_R \approx 0.5)$ when $P(B|R)
~\propto R^{0.7}$.  A crossover in scaling exponents may be useful for
understanding asset bubbles.

\section{A Model}

\begin{figure}
\centering{\includegraphics[width=9cm,angle=-0]{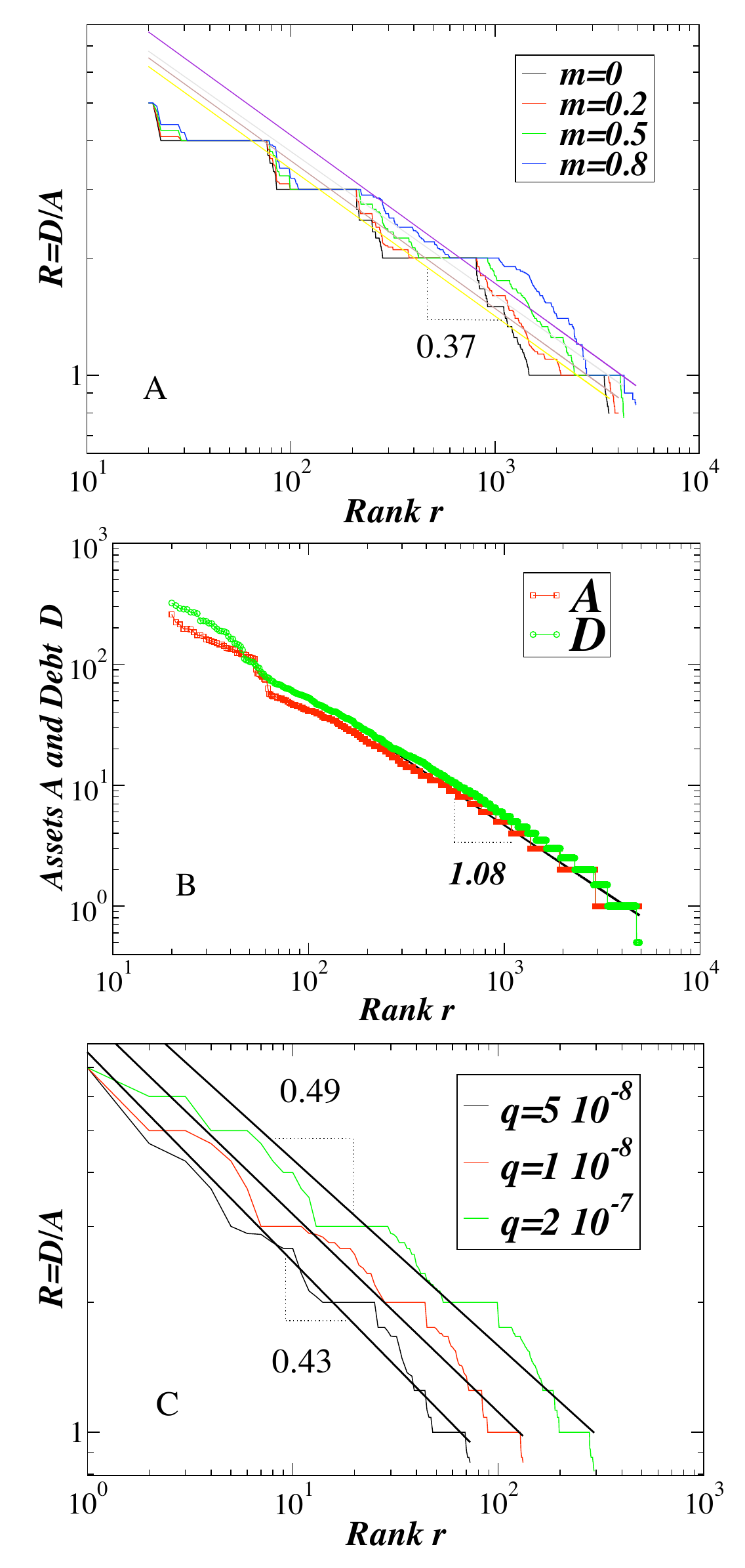}}
\caption{Model results. (a) Zipf plot of debt-to-assets ratio $R$ versus
  rank $r$ for the firms generated by the model (compare with Fig.~4)
  when bankruptcy is not included. In order to understand plateaus in
  the figure, note that both assets and debt in the model exhibit
  integer values.  (b) For each asset and debt the Zipf plot displays a
  power law $R\sim r^{-\zeta}$.  (c) Zipf plot of $R$ versus rank as a
  function of bankruptcy rate parameter $q$. With decreasing $q$ the
  slope increases slightly.}
\label{fig.6}
\end{figure}

\begin{figure}
\centering{\includegraphics[width=9cm,angle=-0]{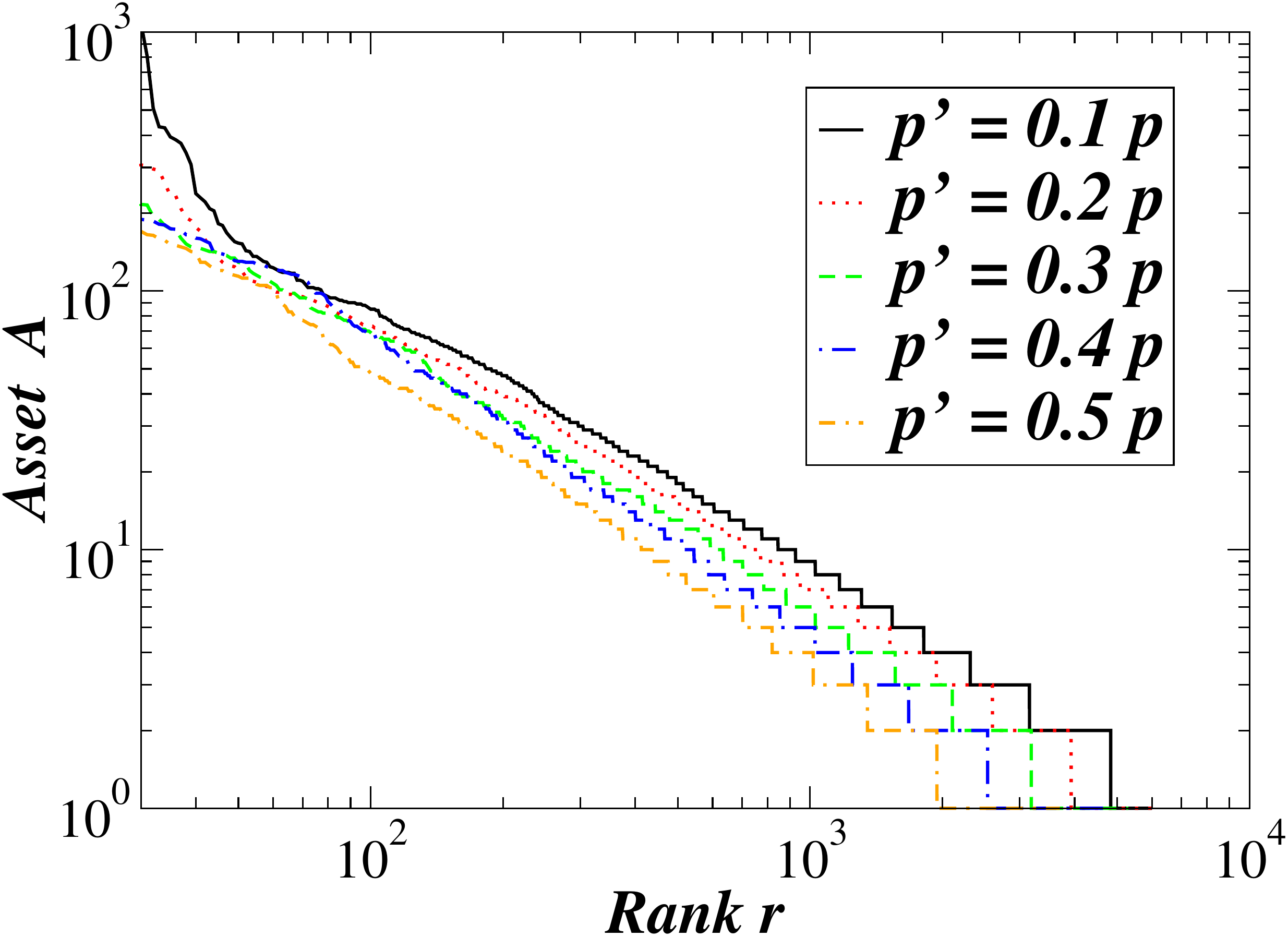}}
\caption{Scale-free persistence in the Simon model with mergers.  Zipf
  plot of value $A_{j}(t)$---the value of assets of the firm
  $j$---versus rank $r$ for the Simon model with a merger parameter $p'$
  representing the probability that a pair of firms will merge.  With
  increasing $p'$, $\zeta$ slowly increases.}
\label{fig.7}
\end{figure}

\begin{figure}
\centering{\includegraphics[width=9cm,angle=-0]{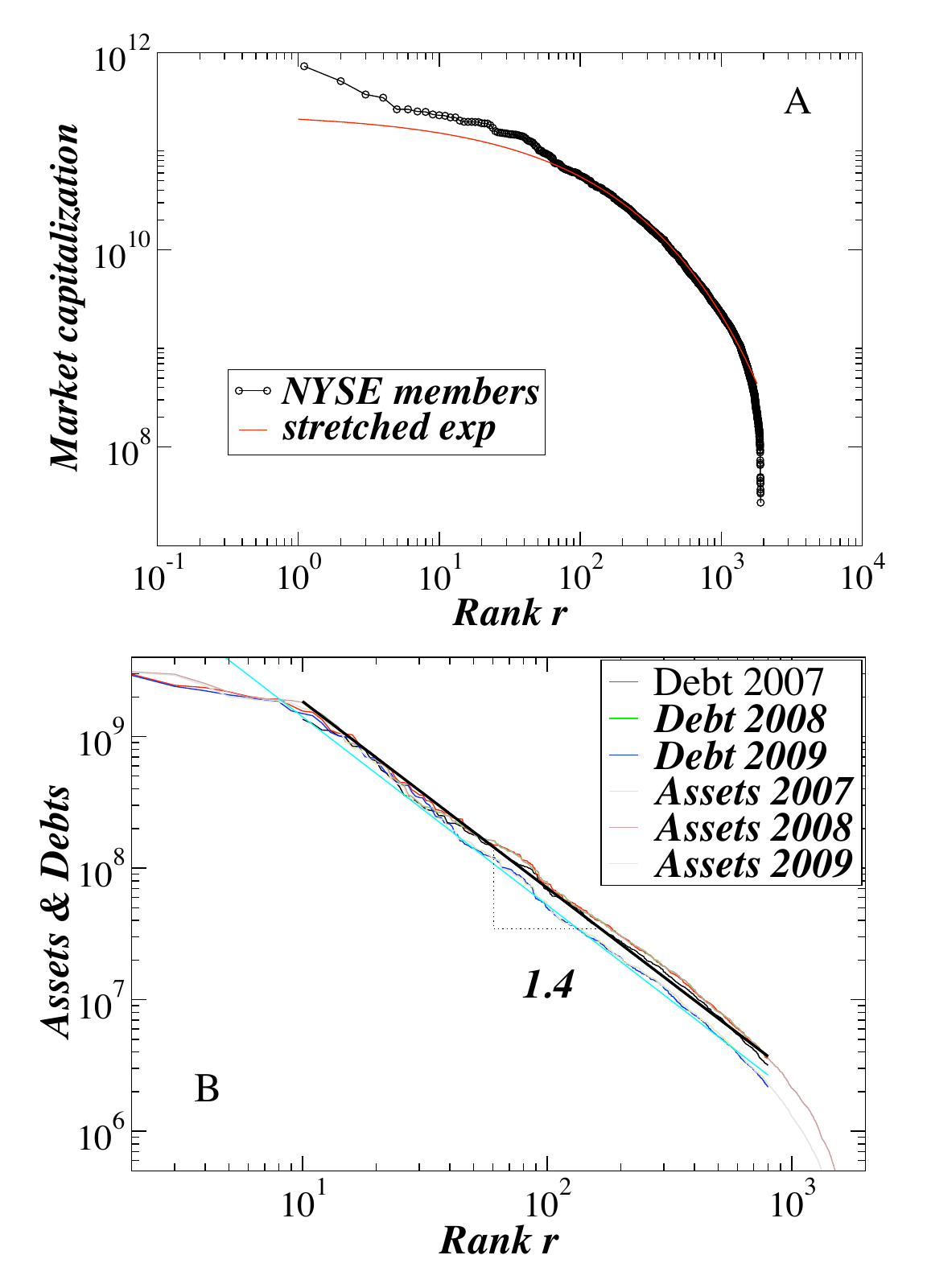}}
\caption{Zipf plot of (a) market capitalization $M$ versus rank $r$ and
  (b) assets and debt versus rank $r$ for the NYSE members for year
  2007. The curve in (a) follows a stretched exponential $exp(-r^{\beta}
  / \tau)$ with $\beta = 0.5$ and $\tau = 45$.}
\label{fig.8}
\end{figure}

Our results complement both the literature on default risk as well as
the literature on firm growth.  According to a study of U.S. firm
dynamics, over 65\% of the 500 largest U.S. firms in 1982 no longer
existed as independent entities by 1996 \cite{Blair1}. To explain how
firms develop, expand and then cease to exist, Jovanovic proposed a
theory of selection where the key is firm efficiency; efficient firms
grow and survive and the inefficient decline and, eventually, fail
\cite{Jovanovic1}.  Many models have been proposed to model default risk
\cite{Duffie, Lando,Merton74,Black76, Long95,Hull95}.  One strain of
that literature \cite{Merton74} develops structural models of credit
risk. In these models, risky debt is modeled within an option-pricing
framework where an underlying asset is the value of company
assets. Bankruptcy occurs endogenously when the value of company assets
is insufficient to cover obligations. In contrast, in reduced form
models \cite{Lando} default is modeled exogenously.

In order to reproduce the Zipf law that holds for bankrupt firms, we
propose a coupled Simon model, an extension of the Simon model used in
the theory of firm growth
\cite{Simon1,Ijiri,Sutton1,Sutton2,pnas1}. Here we couple the evolution
of both asset growth and debt growth through debt acquisition which
depends on a firms assets, and further impose a bankruptcy condition on
a firm's assets and debt values at any given time.

{\it Simon rule for assets.} The economy begins with one firm at the
initial time $t\equiv 1$.  At each step a new firm with initial assets
$A\equiv 1$ is added to the economy. With a probability $p$, a new firm
$i$ is added to the economy as an individual entity at time $t_i$. With
probability $1 - p$, the new firm $i$ is taken over by another firm. The
probability that firm $i$ is taken over by an existing firm $j$ is
proportional to $A_{j}(t)$, the number of units in firm $j$ is equal to
$(1 - p) A_{j}(t) / \Sigma_k A_k(t)$.  Hence, a larger firm is more
likely to acquire a firm than a smaller firm. In this expression, the
index $k$ runs over all of the existing firms at time $t$.  We use the
value $A_{j}(t)$ to be the proxy for the value of assets of the firm
$j$.  Simon found a stationary solution exhibiting power-law scaling,
$P(s>x) \propto s^{-\zeta'}$, with exponent $\zeta' = 1 /(1 -p)$.  For
an estimate of $p$ one can investigate venture data to see how venture
capitalists dispose of their companies.  Even though data suggest $p = 0.5$
(see \cite{Jovanovic4}), we use a much smaller value $p = 0.01$ in oder 
 to reproduce Zipf plot in Eq.~(4).

{\it Simon rule for debt.} When a new firm $i$ is created at time $t_i$,
it is assigned debt $D_i(t_i)= m$, where $0<m<1$. For simplicity we use
a single $m$ value for all firms.  If an existing firm $j$ acquires the
new asset $A_{i}\equiv 1$, then $A_{j}(t)-A_{j}(t-1) =1 $, and debt
$D_{j}(t)-D_{j}(t-1) = m$. Hence, a firm with assets $A_{j}(t)=N$ has
debt $D_{j}(t) =mN$, implying that the debt-to-assets ratio $R=m$ is the
same for all firms.

In order to introduce variation in $R$ ratios across firms, we assume
that at each time $t_i$, a new debt is created in the economy for some
company $j$, so that $D_{j}(t_{i})-D_{j}(t_{i}-1)= 1$.  Hence, for each
time step, there is a new firm receiving debt $D_{i}=m$ in addition to
firm $j$ receiving one unit of debt, where generally $i\neq j$.  The
newly created units of debt are acquired with probability proportional
to $A_{j}(t)$.  Hence, the Simon laws controlling the growth of debt
$D_{j}(t) $ and the growth of assets $A_{j}(t)$ are coupled. In our
model, richer firms become more indebted, but also acquire new firms
with larger probability.

In Fig.~6(a) we perform the numerical simulation of the model by
generating 500,000 Monte-Carlo time steps.  We calculate the Zipf
distribution of the debt-to-assets ratio $R$ for different choices of
$m$. Even though debt and hence $R$ increases with $m$, the slope of the
Zipf plot for $R$ versus rank practically does not depend on the value of
$m$. Unless stated otherwise, in other simulations we set $m=0.5$.

Following \cite{Barabasi99}, we consider the continuous-time version of
our discrete-time model. In this case $D_{j}(t)$ and $ A_{j}(t)$ are
continuous real-valued functions of time. Further, we assume that the
rate at which $D_{j}(t)$ changes in time is proportional to the assets
size $A_{j}(t)$. Hence, following this assumption,
$D_{j}(t)=(1+m)A_{j}(t)$ because of the acquisition of additional debt.
Therefore, since $A_{j}(t) = t/t_{j}$ \cite{Barabasi99}, then
$D_{j}(t)=(1+m)t/t_{j}$. The cumulative probability that a firm has debt
size $D_{j}(t)$ smaller than $D$ is, therefore, $P(D_{j}(t) < D) = P(
t_{j} > (1 + m) t / D)$.  In the Simon model we add new firms at equal
time intervals. Thus, each value $t_{i}$ is realized with a constant
probability $P(t_j) = 1/t $. It follows that
\begin{equation}
P( t_j > (1+m) t / D) =1 - (1+m)t/Dt =  1 - (1+m)/ D.
\label{derivation}
\end{equation}
Since the parameter $t$ cancels out, the same expression we obtain when
$t$ goes to infinity. This is the Zipf law for debt in the case when
there is no possibility of bankruptcy [see Eq.~(3)].

{\it Firm bankruptcy.}  Up to now, debt has been modeled as riskless. We
now introduce bankruptcy into the coupled Simon model.  We assume that
for each firm there is a likelihood of bankruptcy, which depends on the
volatile firm asset value \cite{Merton74}.  In order to be consistent
with our empirical findings, we assume that the firm $j$ that was
created at time $t_j$ files for bankruptcy with probability $q
~R^{0.95}$ [see Eq.~(6)], where $q$ is the bankruptcy rate parameter,
related to $P(B)$ in Eq.~(6). In the hazard model, the hazard rate is
the probability of bankruptcy as of time t, conditional upon having
survived until time $t$ \cite{Shumway}.  In our model, once firm $j$
files for bankruptcy, part of its debt is lost (restructured) and the
firm starts anew with debt equal to $D_{j} = m A_{j}$.  We do not assume
a merger or a liquidation and a firm's probability of failure does not
depend on its age \cite{Shumway}.  Besides bankruptcy, a firm may leave
an industry through merger and voluntary liquidation \cite{Schary}.

Next we perform 500,000 Monte-Carlo time steps for the model with the
possibility of bankruptcy.  Fig.~6(b) presents Zipf distribution for
firm asset and debt values for all of the existing firms.  Each of these
distributions is in agreement with the Zipf law and
Eq.~(\ref{derivation}).  In Fig.~6(c), for the subset of bankrupt
companies, we show the Zipf distribution for $R$ using three different
values of the bankruptcy rate $q$. Note that $q$ is supposed to be
small.  Namely, with $q=10^{-7}$ and with 500,000 time steps
representing one year, $500,000~q$ represents a probability per year
that a company files for bankruptcy during a period of one year,
$\approx 0.05$ in our case.  This should be compared with the average
default rate, $\approx 0.04$, calculated in the period 1985--2007
\cite{Altman07}.  We see that model predictions approximately correspond
to the empirical findings.

Our model can be extended in different ways 
 including mergers between firms. First,  while the
Simon model assumes that at each time increment a new unit is added, we
can assume that the number of new units grows as a power-law
$t^{\theta}$ \cite{Dorogovtsev}.  By using a continuous-time version of
a discrete-time model we obtain $ P(D_{j}(t) < D) = P( t_{j} > (1 + m) t
/ D) = 1 - (\frac{1+m}{D})^{1+\theta}$, where we use $P(t_i > t_0) =
\int_{t_0}^{t} dt t^{\theta} / \int_{0}^{t} dt t^{\theta}$.
Second,  Jovanovic and Rousseau \cite{Jovanovic4} found that mergers
contribute more to firm growth than when a firm takes over a small new
entrant. In order to incorporate mergers into the Simon model, we assume
that at each time $t$, a single merger between a pair of firms occurs
with probability $p'$, where two firms are randomly chosen.
Reference~\cite{Andrade} reported that in more than two thirds of all
mergers since 1973, the Tobin $Q$ value of the acquisition firm exceeded
the Tobin $Q$ value of the target firm, where $Q$ is Tobin 's ratio
similarly defined as $D$ ratio in Eq.~(1).  To this end, we assume that
if $A_j > A_i$ when a merger occurs $A_j = A_j + A_i $ and $ A_i = 0 $.
Thus, the more-rich firm $j$ buys the less-rich firm $i$ resulting in
the elimination of firm $i$ as an individual entity.  In Fig.~7 we show
that the inclusion of mergers does not change the scale free nature of
the Simon model.  In these simulations we use a varying merger
probability $p'$, and $p=0.01$ with 1,000,000 time steps.  With
increasing $p'$, the Zipf exponent $\zeta$ slowly decreases.  Note that
with 1,000,000 time steps if $p' = 0.5 p$, and with $p = 0.01$, then
approximately 5,000 mergers occur.

\section{Discussion and summary}

We demonstrate that Zipf scaling techniques may be useful in analyzing
bankruptcy risk. We find that book values of pre-petition and petition
company assets systematically differ from each other and that the
difference depends upon company size.  Also, the debt-to-assets ratio
$R$ for firms that filed for bankruptcy has a probability distribution
that depends upon the firm size. In analyzing existing (non-bankrupt)
companies we use as a proxy stocks traded at Nasdaq.  We demonstrate
that market capitalization as well as book value of assets, liabilities
and equity for the stocks traded at Nasdaq exhibit Pareto scaling
properties.  This is not a trivial consequence of the scaling
relationship of firm size \cite{Axtell1} since for companies traded at
NYSE we do not find similar power-law scaling for market capitalization
[see Fig.~8(a)] and book value of equity.  However, the book value of
assets and liabilities for NYSE stocks follows a Pareto law with
exponents that are slightly larger than those we find for Nasdaq stocks
[see Fig.~8(b)].  Our results reveal a discrepancy in scaling of market
capitalization and book value of equity obtained from different exchange
markets (e.g., Nasdaq and NYSE).

Using our scaling results, we derive a simple expression for the
conditional bankruptcy probability given $R$.  Importantly, we find that
scaling properties for large $R$ values change during the periods of
significant market turbulence such as the {\tt dot.com} bubble crash
(2001--2003) and the current global crisis where scaling exhibits
significant cross-over properties. Change of scaling exponents may,
therefore, be of significance understanding the asset bubbles.

In order to reproduce our empirical results we model growth of risky
debt and asset values by means of two dependent (coupled) Simon models.
Predictions of the coupled Simon model are consistent with our empirical
findings.


\end{document}